\documentclass[preprint]{aastex}

\begin{document}

\title{The Optical Depth to Reionization as a Probe of
Cosmological and Astrophysical Parameters}
\author{Aparna Venkatesan}
\affil{Department of Astronomy and Astrophysics, \\ 5640
S. Ellis Ave., University of Chicago, \\ Chicago, IL 60637, USA}

\email{aparna@oddjob.uchicago.edu}

\begin{abstract}
Current data of high-redshift absorption-line systems imply that
hydrogen reionization occurred before redshifts of about 5.
Previous works on reionization by the first stars or quasars
have shown that such scenarios are described by a large number
of cosmological and astrophysical parameters.  Here, we adopt a
semi-analytic model of stellar reionization in order to quantify
how the optical depth to reionization depends on such
parameters, and combine this with constraints from the cosmic
microwave background (CMB). We find this approach to be
particularly useful in alleviating the well-known degeneracy in
CMB parameter extraction between the optical depth to
reionization and the amplitude of the primordial power spectrum,
due to the complementary information from the reionization
model.  We also examine translating independent limits on
astrophysical parameters into those on cosmological parameters,
or conversely, how improved determinations of cosmological
parameters will constrain astrophysical unknowns.
\end{abstract}


\section{Introduction}

Observations of the spectra of distant quasars and galaxies have
revealed the absence of a Gunn-Peterson trough, implying that
the intergalactic medium (IGM) was highly ionized by redshifts
of about 5. Since the universe recombined at redshifts, $z$
$\sim$ 1100, the IGM is expected to remain neutral until it is
reionized through the activity of the first luminous sources. At
present, it appears that the reionization of hydrogen occurred
before $z \sim 5$ \citep{ssgunn, hugal, spinrad}, while that of
doubly ionized helium is thought to have occurred before $z \sim
3$ \citep{hog97,reim97}.

The ionizing sources responsible for reionization can be a
variety of astrophysical objects, and much work has been done on
reionization by the first stars \citep{hl97,fuka}, the first
quasars \citep{hl98,valsilk}, proto-galaxies \citep{cenost93,
gn99,gs96,ciardi,mirees99,madrees99}, and related phenomena such
as supernova-driven winds \citep{tse93}, and cosmic rays
\citep{nath93}. Quasars are natural candidates for both H I and
He II reionization, as they are more luminous and provide harder
ionizing radiation relative to stars, and they are seen up to
the highest redshifts of current observations. However, there
are indications of a turnover in the space density of the QSO
population, which apparently decreases beyond $ z \sim 3$. As
this is based on optical surveys, this observed decline is
subject to the effects of any dust obscuration along the line of
sight; however, recent radio observations also appear to
indicate a declining QSO population beyond $z \sim 3$, so that
the comoving emission rate of Lyman continuum photons from QSOs
is deficient by a factor of $\sim$ 4 relative to that required
for reionization (see, e.g., \citet{mad99} and references
therein). If this is real, then QSOs may be less plausible as
sources for H I reionization (see also \citet{rauch97}). The
alternative is to allow QSO formation in collapsing halos from
the outset, and to postulate a large population of faint QSOs at
$z \ga 5$, with the observed turnover being true for bright QSOs
only.

Stellar reionization is attractive for several reasons. The
first stars are expected to form at $z \ga 10$, and are capable
of ionizing hydrogen. Furthermore, they create heavy elements,
and may account for the low (0.01 $Z_\odot$) but persistent
metallicity seen in the Ly-$\alpha$ forest clouds out to $ z
\sim 4$. Finally, the detections of a He II absorption trough in
high-$z$ quasar spectra appear to indicate a soft component to
the UV background \citep{hm96}, consistent with the ionizing
spectrum of stars.

Early work on hydrogen reionization \citep{arwing} described the
appearance of the first luminous sources, about which ionized
bubbles gradually expand into a neutral IGM; eventually such H
II regions overlap and the universe becomes transparent to Lyman
continuum photons. In principle, only one ionizing photon per
neutral atom in the IGM is required, but the effects of
recombinations ensure that for an atom to {\it stay} ionized,
the rate of ionizing photons generated by sources must be
greater than the rate of recombination at that epoch. This is of
particular importance at high $z$'s, when the IGM had greater
density. Just how much more than 1 photon per baryon is needed
is a function of the cosmology and the assumptions of the
reionization treatment; some evolution with redshift is also
expected. A qualitative assessment may be made however; for
example, from arguments of producing the average IGM metallicity
in C IV of about 10$^{-2} Z_\odot$ at $z \la 5$,
\citet{mirees97} arrived at a requirement of 20 ionizing photons
per baryon. For the stellar reionization model that will be
considered in this work, we will see below that about 5 ionizing
photons per baryon are available for H I reionization and prove
sufficient. Not more than a few percent of all baryons need
participate in early star formation for reionization to occur by
$z \sim 5$, though this number may reach values of up to about
15\% (\S 2).

Though reionization by early stars would occur at redshifts well
beyond current observations, it has many distinct consequences
which can be feasibly constrained by current and future
experiments. Some of these include, as mentioned above, the
evolving IGM metallicity and the cosmic ionizing background as
derived from spectral features in high-$z$ absorption-line
systems. High-$z$ reionization will also leave a signature in
the CMB through the Thomson scattering of CMB photons from free
electrons \citep{ssv93,dodel95,tsb94,ts95,hu99,zal97b}. Depending
on the epoch and degree of reionization, we expect an overall
(somewhat scale-dependent) damping of primary temperature
anisotropies in the CMB, the generation of new temperature
anisotropies on the appropriate scales through the effects of
second-order processes and the degree of inhomogeneity in the
reionization process \citep{grhu,ksd98}, and finally, the
creation of a new polarization signal, as the process of Thomson
scattering introduces some degree of polarization even for
incident radiation that is unpolarized. Scattering from the
ionized IGM, or the reprocessing of starlight into far-infrared
wavelengths by dust formed from early supernovae (SNe), will
also cause the CMB to undergo some spectral distortion
\citep{lh97}; this can be measured experimentally through the
Compton $y$-parameter. These and other observational signatures
that have the potential to constrain the epoch, and hence
possibly the source, of reionization have been examined in the
literature (see \citet{hk99} for a summary).

A model of reionization is therefore, in principle, eminently
testable. Current detections of the first Doppler peak in the
CMB's temperature anisotropies limit the total optical depth to
electron scattering, $\tau_{e}$, such as may arise from
reionization, to be $\tau_e \la 1$ \citep{sswhite}. Future
experiments such as the {\it Next Generation Space Telescope
(NGST)} or the {\it Space Infrared Telescope Facility} ($SIRTF$)
may detect the high-$z$ sites of reionizing sources (see, e.g.,
\citet{hlngst}), or at least exclude currently viable
candidates, while upcoming CMB experiments such as $MAP$ or the
{\it PLANCK} surveyor can measure $\tau_{e}$ to very high
accuracies by combining information from temperature
anisotropies and polarization in the CMB.

The optimistic prospects for testing reionization and the
increasing multiwavelength view of the high-$z$ universe have
generated a large body of work on reionization models in the
last few years, whose techniques fall broadly into numerical
\citep{gn99,co99,go97,cenost93} or semi-analytic methods \citep{hl97,
hl98,tsb94,valsilk}. The former have the advantage of being able
to track the details of radiative transfer, incorporating the
clumpiness of the IGM and the essentially non-uniform
development of ionizing sources, and, perhaps most importantly,
describing the {\it process} of reionization in a quantitative
fashion. The advantage of semi-analytic approaches is their
inherent flexibility and ability to probe the parameter space of
a reionization model at will, which is of value given the many
input cosmological and astrophysical parameters involved.

For astrophysical sources, the process of reionization is
strongly related to the evolution of structure in the universe,
and could result in feedback effects for subsequent object
formation (see, e.g., \citet{ciardi}). Of the current theories of
structure formation, variants of the standard cold dark matter
(sCDM) model are considered to be relatively successful at
describing the observed universe. This picture postulates a
critical density universe, with cold dark matter dominating the
matter content; structures, made up of baryons and CDM,
originated in primordial adiabatic fluctuations and evolved
subsequently through gravitational instability. Current
modifications to this paradigm include, e.g., the addition of a
cosmological constant.

The sCDM model, and its variants, are described by a set of
parameters which characterize the primordial power spectrum of
fluctuations, the cosmology of the universe, and quantities
related to primordial nucleosynthesis.  At present, the
extraction of such parameters from observations has been made
feasible by the quality of data from large-scale structure
surveys, from cosmic velocity flows \citep{zeh99}, from Type Ia
SNe \citep{teg99}, and from current and projected future data
from the CMB \citep{zal97a, eht99}.  Typically, a 9--13
parameter set describes the adiabatic CDM model, and can be
solved for given the data. One of these parameters is $\tau_e$,
which is by nature somewhat unique, in that it is the only
quantity that is not set purely by the physics prior to the
first few minutes after the Big Bang. Thus it can potentially
provide clues on post-recombination astrophysics, assuming that
the other (cosmological) parameters which also affect $\tau_e$
are comparatively well-constrained.

Several of the semi-analytic works on reionization mentioned
above have explored the effects of varying model parameters on
$\tau_e$. Other authors have performed CMB analyses that have
revealed inherent degeneracies in constraining specific
combinations of parameters, e.g., $\tau_e$ and the amplitude of
the primordial power spectrum, $A$ (see, e.g., \citet{zal97a,
eht99}).  In this paper, we examine the results of
cross-constraining the range in a cosmological parameter, given
the allowed band in $\tau_e$ from a reionization model due to
astrophysical parameter uncertainty, with the permitted range
from CMB observations. Specifically, we find that for the
combination $\tau_e$--$A$, the well-known degeneracy in their
effects on the CMB can be broken when used in combination with
the constraints from a reionization model. As $\tau_e$ depends
on both cosmological and astrophysical parameters, however, such
an analysis can be extended to mutual constraints involving
these two independent classes of parameters, by eliminating
$\tau_e$. The advantage of this is that, given a model of
structure formation and a reasonable framework describing
reionization, as well as the data from the CMB, we can use known
astrophysics to further constrain cosmology and place tighter
limits on cosmological parameters, even those that will be
determined to high accuracies by future experiments. Conversely,
a well-determined cosmological parameter can be used to
constrain the astrophysics of ill-known details of early star
formation. This will, at the least, be a powerful test of the
cosmology, if our understanding of reionization is reasonably
correct; the additional hope is that this will prove to be an
alternative way of constraining the activity of the first stars.

The plan of the paper is as follows. In \S 2, we outline the
stellar reionization scenario that we consider, and set up a
parameter set that describes reionization. In \S 3, we review
the standard methodology related to parameter extraction from
the CMB, and incorporate the parameter set from \S 2 into this
formalism. In \S 4, we present our results, and show the
combined constraints from a reionization model and the CMB. We
present our conclusions in \S 5.

\newpage
\section{The Stellar Reionization Model}

We assume a sCDM primordial power spectrum of density
fluctuations, given by $P(k) = A k^n T^2 (k)$, where $A$ and $n$
are, respectively, the amplitude and index of the power
spectrum, and the matter transfer function $T(k)$ is taken to be
the form given in \citet{bbks}, with the baryon correction as
given by \citet{pdod}.  Here, we evaluate the power spectrum
normalization $A$ through the value of the rms density contrast
over spheres of radius 8 $h^{-1}$ Mpc today, $\sigma_8$.  The
cosmology of our model is also described by $\Omega_0$, the
cosmological density parameter, $\Omega_b$, the density
parameter of baryons, and $h$, the Hubble constant in units of
100 km s$^{-1}$ Mpc$^{-1}$. We will assume that $\Omega_0 = 1$
throughout this work, and thus will not include it in the
parameter set to be varied in what follows. We assume that there
are no tensor contributions to $P(k)$, and set the cosmological
constant to be zero. Thus, our cosmological parameter set is
[$A$, $n$, $\Omega_b$, $h$].

The reionization of the universe by the first generations of
stars is described by the model developed in \citet{hl97}
(henceforth HL97), with the minor modifications described
below. Briefly, the fraction of baryons in collapsed dark matter
halos, $F_B$, is followed using the Press-Schechter formulation;
of these baryons, a fraction $f_\star$ cool and form stars in a
Scalo initial mass function (IMF).  A fraction $f_{esc}$ of the
generated ionizing photons is assumed to escape from the host
object and propagate isotropically into the IGM. One can then
solve for the size of the ionized regions associated with each
such star-forming cloud, which, when integrated over all haloes,
yields at each $z$ the average ionization fraction of the
universe, given by the filling factor of ionized hydrogen by
volume ($F_{H II}$). We assume that the IGM is homogeneous, in
which case the ionized region created by each source can be
taken to be spheres of radius $r_i$. Reionization is defined to
occur when $F_{H II}$ = 1. The total optical depth for electron
scattering, $\tau_{reion}$, to the reionization redshift
$z_{reion}$, is given by integrating the product of the electron
density, the ionization fraction, and the Thomson cross section
along the line-of-sight from the present to $z_{reion}$. The
cosmology of the universe enters $\tau_{reion}$ through the
first two terms of the integrand, and also through the path
length of the photons last scattered at $z_{reion}$.

Our adopted model is summarized by the following equations for
an $\Omega_0 = 1$ universe:

\begin{equation}
F_B(z) = \rm{erfc}(\frac{\delta_c}{\sqrt 2 \sigma(R,z)})  
\end{equation}
\begin{equation}
F_{H II}(z) = \rho_B(z) \int_{z_\star}^z dz_{on}
 \frac{dF_B}{dz}(z_{on}) \; (\frac {4 \pi}{3 M} r_i^3(z_{on},z)) 
\end{equation}
\begin{equation}
\tau_{reion} = 0.053 \; \Omega_b \; h \int_0^{z_{reion}} dz
\; \sqrt{1 + z} \; (1 - f_\star F_B(z)) F_{H II}(z)
\end{equation}

The critical overdensity $\delta_c \equiv 1.686$, $\sigma(R,z)$
is evaluated with a spherical top-hat window function over a
scale $R \propto M_C$, where $M_C$ is the minimum halo mass that
collapses at a given redshift. While a natural choice for $M_C$
is the baryonic Jeans mass, given by $\sim 10^6 M_\odot [(1 +
z)/100]^{1.5}$, this assumes that collapsing halos at a given
mass scale are equivalent to star-forming halos. However,
several authors \citep{teg97, hrl96} have argued for a higher
value of $M_C$, based on the requirement of an effective coolant
for baryons in a halo to fragment into stars. The picture is as
follows: the very first stars form from metal-free gas and cool
through primordial molecular hydrogen. The universe at that
epoch is transparent to photons in the energy range 11.2--13.6
eV, but is opaque to more energetic photons. This initial trace
level of stellar activity easily photodissociates the remaining
H$_2$ in the universe (whose abundance relative to H I is very
low), well before the H I ionizing flux has built to values
sufficient for reionization (see HL97 and references
therein). Subsequent halo formation continues but star formation
halts as there is no coolant available, and can only resume when
halos more massive than $10^8 M_\odot [(1 + z)/10]^{-1.5}$
collapse, which can utilize line cooling by atomic H. We set
$M_C$ to be this higher value, with the understanding that $F_B$
now represents the fraction of all baryons that are in {\it
star-forming} halos. Finally, $z_\star$ is the earliest redshift
at which the first stars can form, and here we set it to be
100. Prior to this, the temperature of the IGM is coupled to
that of the CMB, and the CMB photons are still energetic enough
to photodestroy H$^-$, thus preventing the formation of H$_2$
through the H$^-$ channel, an important cooling mechanism for
structures at these high redshifts to fragment into stars.

The evolution of an individual ionization front is characterized
by the ionization radius $r_i$, and for a time-dependent source
luminosity, can be solved for through a differential equation as
in HL97, where the rate of emission of ionizing photons from a
stellar population of metallicity $Z = 10^{-4} Z_\odot$ was
constant for about 2 million years before declining with the
death of the massive stars in the IMF. In this work, we use the
analytic solution from \citet{sg87} (SG87 hereafter) for the
evolution of $r_i$ in an expanding universe in units of the
Str\"{o}mgren radius $r_s$. The Str\"{o}mgren radius represents
the equilibrium reached in the IGM between a source's ionizing
photon rate and the IGM's recombination rate; $r_s$ increases
with decreasing $z$, or equivalently, with decreasing average
IGM density. The maximum value that $r_i$ can have is $r_s$;
only sources at very high redshifts ($\sim 100$) have ionized
regions that fill their Str\"{o}mgren spheres [SG87]. Note that
as the SG87 solution does not account for time-varying sources,
we expect $\tau_{reion}$ to be overestimated (reionization
occurs earlier) compared to HL97, but as we will see in the next
section, this is a very slight effect. Thus, in this work, $r_i
= (r_i/r_s)_{SG87} \, r_s$, where $r_s^3 = 3 S(0)/(4 \pi
\alpha_B n_H^2(z))$, $\alpha_B = 2.6 \times 10^{-13}$ cm$^3$
s$^{-1}$, and the initial emission rate of ionizing photons
leaving the host object is $S(0)$. $F_{H II}(z)$ (see eq. [2])
is determined at each redshift $z$ by integrating over the
product of the rate of new halos that formed stars at a turn-on
redshift $z_{on}$ (where $z< z_{on} < z_\star$), and the ionized
volume per unit mass associated with such objects, for a source
mass $M$. A detailed treatment of the evolution of $F_B$ and
$F_{H II}$ with $z$, given various choices of input parameters,
may be found in HL97. For their standard model, $F_B$ rises
rapidly from values of $\sim$ 10$^{-3}$ at $z \sim 20$ to about
0.1 at $z \sim 10$; during this period, $F_{H II}$ rises steeply
from about 10$^{-4}$ ($z \sim 32$) to unity at $z \sim 18$, so
that reionization occurs relatively quickly with the growth of
$F_B$.

We now see that $\tau_{reion}$ is a function of several
cosmological and astrophysical parameters. The astrophysical
parameter set is [$f_\star, M_C, S(0)$]. $S(0)$ is itself a
function of several variables: the choice of the IMF, the
metallicity $Z$ of the progenitor stars, $f_{esc}$, $f_\star$,
and the halo's mass. The last two factors account for the
fraction of star-forming baryons in each halo, while $f_{esc}$
represents the loss of ionizing photons to the host cloud before
reaching the IGM. Let us now consider the first two
variables. Though there have been some arguments for the IMF to
be biased towards high-mass stars in the early universe
\citep{lar98}, the details of the nature of star formation under
those conditions are still not well understood. In the absence
of a convincing theory of primordial star formation, the most
reasonable assumption is to take a present-day IMF and calculate
the luminosity expected from metal-poor stars. As reionization
is affected primarily by the massive stars in any IMF, an IMF in
the past biased towards high-mass stars would still have the
same emission spectrum of ionizing photons, while one dominated
by low-mass stars is unlikely to reionize the universe by $z
\sim 5$. We therefore take S(0) to be as given in HL97 for $Z =
10^{-4} Z_\odot$ stars from standard stellar evolutionary
models; it includes the ionizing radiation from stars only, and
is steady at $ f_\star f_{esc} 10^{46}$ photons s$^{-1}
M_\odot^{-1}$ for about 2 million years before declining
rapidly, which is consistent with the value in, e.g.,
\citet{ciardi}.  As a rough estimate, this translates to $\sim$ 5
ionizing photons per baryon in the universe, for $f_\star$ =
0.05, $f_{esc}$ = 0.2.

The ionizing photon contribution from SNe is relatively small
(HL97), but most of the {\it mass} assigned to forming stars is
eventually returned to the IGM. Therefore, the second factor on
the RHS of equation (3) for $\tau_{reion}$ should have an extra
contribution $f_{SN} f_\star F_B$ to account for the extra
baryons that are available for new star formation, where
$f_{SN}$ is an IMF-averaged fraction of the progenitor mass that
is expelled into the IGM at the end of the star's life. We
expect between $\sim$ 50--95\% of the parent star's mass as
ejecta, until a point is reached (for stellar masses ranging
from 50--100 $M_\odot$) where the entire star collapses into a
black hole. For the low values of $f_\star$ that will be
considered here, the product $f_\star f_{SN}$ will be a small
correction and may be ignored. Moreover, the mass of the ejecta
from a dying star depends sensitively on the stellar
metallicity, with low-$Z$ stars having higher remnant masses and
less ejected material relative to solar-$Z$ stars
\citep{ww95}. Thus, $f_{SN}$ is likely to be highly variable,
both spatially and with $z$ due to the evolving metallicity of
subsequent generations of stars, and is more appropriately
modelled in a simulation rather than in a semi-analytic
model. The calculated values of $\tau_{reion}$ here may be taken
as a lower limit.

This leaves the astrophysical parameters, $f_\star, f_{esc}$,
and $M_C$. We note that in most semi-analytic models, $f_\star$
is set by the choice of $M_C$, as the stellar $Z$-output,
particularly in $^{12}$C, is combined with the evolution of
$F_B$ to produce the observationally detected average carbon
abundance of 0.01 $Z_\odot$ in the Ly-$\alpha$ forest clouds at
$z \sim 3$ \citep{songcow}. Thus, $f_\star$ and $M_C$ are not
independent of each other if we choose the above normalization;
for the HL97 choice of $M_C$, $f_\star = 0.13$. This is the
maximum value that $f_\star$ can have from arguments of avoiding
IGM over-enrichment; given the approximately order-of-magnitude
scatter in the average metal abundance of the Lyman-$\alpha$
clouds, and that one need not require the reionizing stars to
solely account for $Z_{IGM}$, $f_\star$ may be smaller than
$\sim$ 0.15.

\newpage
As an aside for the interested reader, we note here two
drawbacks of normalizing $f_\star$ via $^{12}$C. One is that the
massive stars in the IMF ($\ga 10 M_\odot$) are the ones
relevant for reionization, while $^{12}$C is produced dominantly
by the intermediate-mass stars (2--8 $M_\odot$). Thus, if the
IMF was different in the past, the carbon abundance in the
Ly-$\alpha$ forest does {\it not} constrain the massive or
reionizing stellar activity in early halos. The second point to
note is that the pause in star formation caused by the initial
dissociation of H$_2$ led to the choice of $M_C$, as in
HL97. This minimum halo mass corresponds to objects of virial
temperature of $\sim 10^4$ K. At this temperature, the host
object is immune to photoionization heating (as pointed out in
HL97), and so if outflows are desired to expel the generated
metals into the IGM (i.e, the Ly-$\alpha$ forest), the
mechanical energy of SNe must be invoked. Again, the massive
reionizing stars end their lives as Type II SNe an order of
magnitude in time before their intermediate-mass
carbon-producing compatriots. As all the mechanical input lies
with Type II SNe, the question arises of how the carbon,
produced significantly later, leaves the host halo to mix with
the IGM. Furthermore, Type II SNe occur on much more predictable
timescales, i.e. immediately following the progenitor's death,
than do Type Ia SNe (3--10 Gyr), and there is not much more than
a wheeze to be had from the deaths of intermediate-mass stars as
planetary nebulae.

Having voiced these objections, we point out that while the
$^{12}$C connection as made above between $M_C$ and $f_\star$ is
not ideal, postulating a general relation between these two
variables is not {\it ad hoc}. The value of $M_C$ does
intrinsically determine the stellar history, metallicity and
luminosity evolution of the universe; the high value of $M_C$ in
HL97 and other works is physically well-motivated by the
necessary step of having an available coolant to aid star
formation. We proceed to set $M_C$ = $10^8 M_\odot [(1 +
z)/10]^{-1.5}$ for the semi-analytic treatment here, and now
narrow our astrophysical parameter set to [$f_{esc}, f_\star$].

We end here by addressing some of the issues that are not
accounted for in this work. The IGM is assumed to be
homogeneous, but clearly some clumpiness will develop in the IGM
from the growth of initial density inhomogeneities, and the
assumption of the average ionized fraction at a given redshift
being equal to the H II filling factor will eventually break
down. However, this appears to be a relevant effect only at
``late'' times ($z \la 10$), when the fraction of baryons in
collapsed structures becomes significant \citep{go97}, or for
baryon-dominated universes [SG87]. Therefore, we will assume
that the clumping factor is unity (homogeneous IGM) for the rest
of this work. We have also neglected corrections from doubly
ionized helium, which is not problematic as the spectrum of
photons produced by stars is softer than that from quasars, and
is more relevant for H I rather than He II reionization (see,
however, \citet{tumshull} on the helium-ionizing spectrum from
zero-metallicity stars). We have set $F_{He II} = F_{H II}$, but
this introduces an error of not more than a few percent
\citep{ts95}. Finally, we have not included the effects of bias
in the normalization of the matter power spectrum, i.e., we
assume that light traces the underlying mass distribution.

\section{Constraints From The Microwave
Background} 

As discussed in the introduction, signatures from reionization
are expected in the CMB; an accurate measurement of
$\tau_{reion}$ or the detection of post-recombination features
in the CMB anisotropies have the power to constrain the
reionization epoch and the nature of the sources through the
angular scale $\theta (\propto l^{-1}$) on which they affected
the CMB. Here, $l$ is defined from expanding the angular power
spectrum of the CMB in terms of its multipole moments $C_l$ and
Legendre polynomials:

\begin{equation}
C(\theta) \equiv \sum_{l = 2}^{\infty} \frac{(2l + 1)}{4 \pi}
C_l P_l(cos \, \theta) 
\end{equation}

The effect of $\tau_{reion}$ is to introduce an overall damping
of the temperature $C_l$s (HL97, and references therein), except
at the largest scales. As discussed by, e.g., \citet{zal97a},
this is practically indistinguishable from the generally reduced
values of $C_l$ expected from simply having a lower amplitude,
$A$, of the primordial power spectrum; the difference between
these two effects at the smallest $l$'s is obscured by cosmic
variance.  While the amount of the damping due to $\tau_{reion}$
is $l$-dependent and can potentially break this degeneracy, the
accuracy to which $\tau_{reion}$ can be estimated from
temperature anisotropy maps alone is not sufficient to
distinguish between these two different effects \citep{zal97b}.

When combined with the polarization data from the CMB however,
$\tau_{reion}$ can be constrained with far greater
precision. Linear polarization is generated by the primary
temperature quadrupole anisotropy photons scattering off the
free electrons in the reionized IGM, and is a relatively clean
probe of the epoch of reionization \citep{zal97b}. The
polarization signal is expected at low levels compared to that
from temperature anisotropies, and may prove difficult to
measure, especially for low optical depths. Nevertheless,
$\tau_{reion}$ can be detected, in principle, by future
experiments to within 1-sigma errors of, e.g., 0.69 (0.022)
without (with) polarization information for $MAP$, and 0.59
(0.004) correspondingly for $PLANCK$ \citep{eht99}.

We wish to combine the constraints on cosmological or
astrophysical parameters from a reionization scenario with those
from the CMB; in order to do this for the latter, we follow the
standard prescription as outlined in, e.g., \citet{jung96} and
\citet{knox95}. We assume Gaussian initial perturbations, and
that the multipole moments $C_l$ are determined by a ``true''
set of $N$ theoretical parameters [$P_N$]. If we define the
likelihood function $\mathcal{L}$ of observing a set of $C_l$s,
given $P_N$, then the behavior of $\mathcal{L}$ near its maximum
can be quantified in terms of the Fisher information matrix,
whose elements are given by the second derivative of the
logarithm of $\mathcal{L}$ with respect to pairs of parameters
in $P_N$. The Fisher matrix then represents the accuracy with
which $P_N$ can be estimated from a given data set, here the
CMB's experimentally measured $C_l$s. Further assuming that
$\mathcal{L}$ has a Gaussian form near its maximum, the Fisher
matrix is given by:

\begin{equation}
 F_{ij} = \sum_{l = 2}^{\infty} \frac{1}{\sigma_l^2} \; \left[
\frac{\partial C_l (P_N)}{\partial P_i} \frac{\partial
C_l (P_N)}{\partial P_j} \right], \; \; 1 \leq i,j \leq N
\end{equation}

where $\sigma_l$ is a measure of how the observed $C_l$s are
distributed about the mean value of the true $C_l (P_N)$s. We
assume that $\sigma_l$ is cosmic-variance-limited, and ignore
terms arising from the instrumental noise associated with an
experiment and from any foregrounds. For a sky fraction $f_{\rm
sky}$ that has been mapped, $\sigma_l$ can be approximated by:
\begin{eqnarray}
\sigma_l  = & \sqrt{\frac{2}{(2l +1) f_{\rm sky}}} \; C_l (P_N)
& (l \leq l_{max})  \nonumber \\
 & \infty & (l > l_{max}) 
\end{eqnarray}

We will consider two cases here: ($l_{max}$ = 400, $f_{\rm sky}$
= 0.01), which is roughly representative of data from current
CMB experiments, and ($l_{max}$ = 3000, $f_{\rm sky}$ = 0.8) for
the data expected from $PLANCK$. As we neglect any experimental
or systematic effects, the power of the $C_l$s to constrain
$P_N$, as presented here, is the ``best possible'' case. Note
also that the above formulae are valid when only the temperature
information from the CMB is used. More general expressions for
the case of including polarization data may be found in, e.g.,
\citet{zal97a}.

The derivatives of the $C_l$s with respect to $P_N$ were
computed for each parameter using two-sided derivatives with
step sizes being chosen so that the value of this derivative
remained stable (see, e.g., \citet{eht99}, Appendix B.1). The
$C_l$s themselves for a given parameter set were found using the
publicly available CMBFAST (v.
2.4.1)\footnote{http://www.sns.ias.edu/$\sim$matiasz/CMBFAST/cmbfast.html}.
Note that the parameter set describing the reionization model is
[$A$, $\Omega_b$, $h$, $n$, $f_{esc}$, $f_\star$], which yields
$\tau_{reion}$, whereas the CMB data can determine the
cosmological parameters and $\tau_{reion}$, or equivalently,
[$P_{\rm cosmo}, \tau_{reion} (P_{\rm cosmo}, P_{\rm
astro})$]. Therefore, when we specify $\tau$ (e.g, to CMBFAST),
the cosmological and the astrophysical parameter sets ($P_{\rm
cosmo}, P_{\rm astro}$) are no longer independent, but are
related through the reionization model, and the $C_l$
derivatives become:

\begin{equation}
\frac{\partial C_l}{\partial P_{\rm cosmo}} = \; \left. \frac{\partial
C_l}{\partial P_{\rm cosmo}} \right|_{\tau} \; + \; \;
\left. \frac{\partial C_l}{\partial \tau} \right|_{P_{\rm
cosmo}} \frac{\partial \tau}{\partial P_{\rm cosmo}} 
\end{equation}
\begin{equation}
\frac{\partial C_l}{\partial P_{\rm astro}} = \; \frac{\partial
C_l}{\partial \tau}  \frac{\partial \tau}{\partial P_{\rm
astro}} 
\end{equation}

Once the Fisher matrix $F_{ij}$ has been constructed, it can be
inverted to give the covariance matrix $\mathcal{C}$ between the
parameters $P_N$; $\mathcal{C}_{ii}$ represents the minimum
variance in the estimate of $P_i$.  Any 2 $\times$ 2 submatrix
of $\mathcal{C}$ can then be extracted, giving the ellipse
equation for the joint confidence region in the 2-parameter
subspace of interest,

\begin{equation}
{\bf [P - P_N]} \bullet (\mathcal{C}_{2 \times 2})^{-1}
\bullet {\bf [P - P_N]} = \Delta, 
\end{equation}

where $\Delta$ is set throughout this work to be at the 68\%
confidence level.

\section{Results}

We present our results here from combining the reionization
model (\S 2) and the constraints from the CMB (\S 3). This
analysis assumes that the density perturbation spectrum at the
CMB and structure formation scales is described by the same
power law. For the choice of parameters in HL97, where $f_\star$
= 0.13 and $f_{esc}$ = $f_{esc}(z)$, we obtain $\tau_{reion}$ =
0.0734, or $z_{reion} \sim 18.4$, which only slightly exceeds
the HL97 value of $\tau_{reion}$ = 0.07. Henceforth, we will
refer to $\tau_{reion}$ as $\tau$ for convenience, noting that
$\tau$ is always evaluated to the reionization epoch in this
work.  We define our standard model (SM), fixing $\Omega_0 = 1$,
as given by $A(\sigma_8 = 0.7$) = 1.55 $\times 10^6$, $\Omega_b$
= 0.05, $h$ = 0.5, $n$ = 1.0, $f_{esc}$ = 0.2, $f_\star$ = 0.05,
and $\tau$ = 0.0573, with reionization occurring at $z \sim
15.5$.

\begin{figure}[htb]
\begin{center}
\includegraphics[height=3.5in]{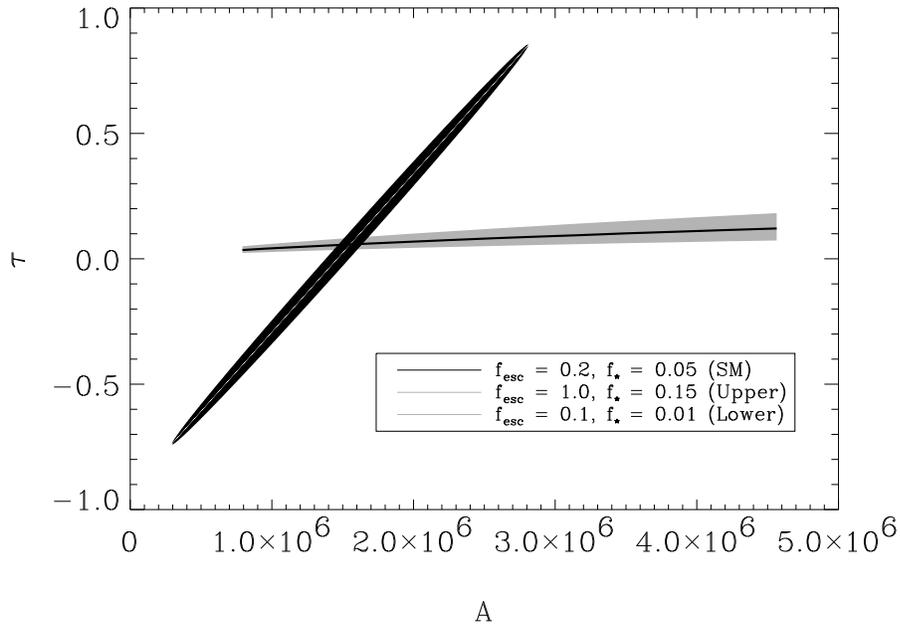}
\end{center}
\caption{Combined constraints in the $\tau$--$A$ plane from the
reionization model and current CMB data. The standard model (SM)
is given by the solid line, the shaded band represents the
astrophysical uncertainty in the reionization model, given the
allowed ranges for ($f_{esc}, f_\star$), and the ellipse is the
1-$\sigma$ joint confidence region from current CMB data.}
\label{fig:taua}
\end{figure}
\begin{figure}[htb]
\begin{center}
\includegraphics[height=3.5in]{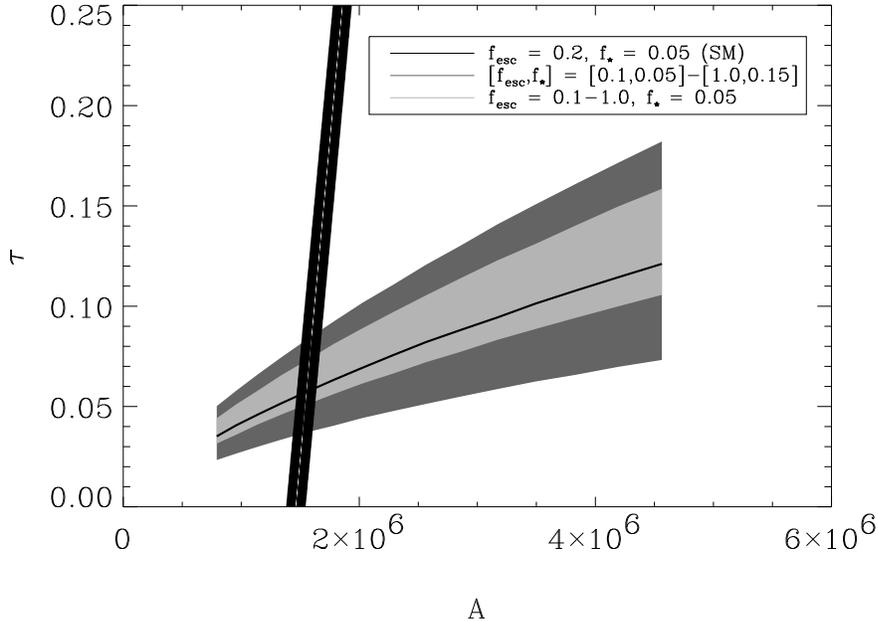}
\end{center}
\caption{Magnified version of Figure \ref{fig:taua}; the
additional light nested band represents the uncertainty in the
value of $f_{esc}$ alone.}
\label{fig:tauamag}
\end{figure}
\begin{figure}[htb]
\begin{center}
\includegraphics[height=3.5in]{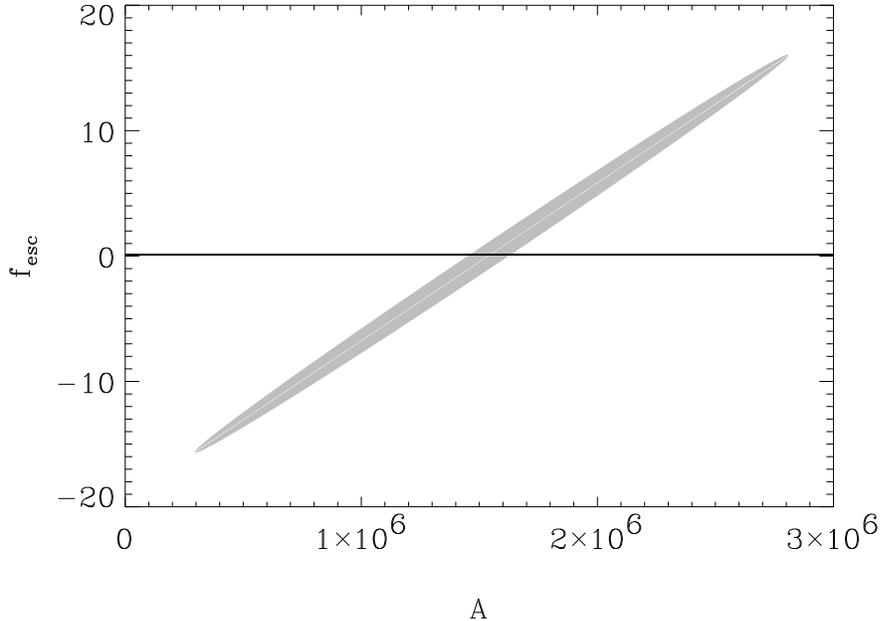}
\end{center}
\caption{Constraint from current CMB data in the $f_{esc}$--$A$
plane, as extended from Figure \ref{fig:tauamag}; the horizontal
band is the permitted range of 0.1--0.2 for $f_{esc}$ from DSF.}
\label{fig:fesca}
\end{figure}

As a simple example, we begin with the $\tau$--$A$ plane, shown
in Figure \ref{fig:taua}, where we isolate the dependence of
$\tau$ on $A$, keeping all the other parameters in the SM
fixed. The range of $\sigma_8$ is $\sim$ 0.5--0.8, from the
large-scale distribution of clusters of galaxies (see, e.g.,
\citet{bunn97}, and references therein), but is 1.2 when
normalized to $COBE$ for the SM choice of cosmological
parameters. As an illustrative range for the plots, we normalize
$A$ for $\sigma_8$ = 0.5--1.2. The solid line represents the SM,
while the light shaded region represents the uncertainty due to
the astrophysics of the reionization model (or AS for
astrophysical slop), with [$f_{esc}$, $f_\star$] = [1.0 (0.1),
0.15 (0.01)] for the upper (lower) envelope. The range for
$f_\star$ (0.01--0.15) is set as follows: the lower limit comes
from numerical simulations of star formation (see, e.g.,
\citet{ciardi}, and references therein), while the upper limit
corresponds approximately to that in HL97 and \citet{hl98}. The
value of $f_{esc}$ has been estimated through a number of
theoretical and observational methods (see \citet{wood99}, and
references therein). Here, we take the range for $f_{esc}$ to be
0.1--1.0, the lower limit coming from \citet{dsf99} (henceforth
DSF), who modelled the escape fraction of ionizing photons from
OB stellar associations in the H I disk of the Milky Way, and
found that for a coeval star formation history, $f_{esc}$ = 0.15
$\pm 0.05$. Our choice of this limit from DSF is motivated by
the similarity of their model's luminosity history to that in
HL97; as noted above, there are alternate values for $f_{esc}$
in the literature for a variety of astrophysical
environments. We will use both the full range for $f_{esc}$ and
the more narrow DSF band in later plots.

To combine this with the CMB constraint, a shaded ellipse
representing the 2-parameter 68\% joint confidence region is
overplotted, assuming that the true model describing the
universe is given by the SM and [$\tau, A$] = [0.057,
$A(\sigma_8 = 0.7$)]. This ellipse is narrow enough that we show
a magnified version of Figure \ref{fig:taua} in Figure
\ref{fig:tauamag}, with an additional lighter band, nested
within the AS band, showing the effect of varying only $f_{esc}$
while fixing $f_\star$ to its SM value. We see that the CMB does
not really constrain $\tau$ or $A$ separately at all, a
near-degeneracy that was expected from the discussion in the
previous section.  However, the combination of the CMB
confidence region and the AS band is much more constraining:
this translates to a 1-$\sigma$ error of about 0.02 for
$\sigma_8$, which is noticeably better than the corresponding
value of $\sim$ 0.2 from the CMB ellipse alone.

We now extend Figure \ref{fig:tauamag} to connect, through
$\tau$, two {\it a priori} independent parameters, $f_{esc}$ and
$A$, shown in Figure \ref{fig:fesca}. The purpose of this plot
is to probe the potential of cosmology and the astrophysics of
reionization to constrain each other, given a reionization
model. The ellipse here, as in Figure \ref{fig:taua}, reveals
the inherent degeneracy between $\tau$-related quantities and
$A$ through the long narrow ellipse. We now overplot the DSF
permitted band for $f_{esc}$ (0.1--0.2); clearly even this
approximate range in $f_{esc}$ considerably narrows the allowed
range in $A$. We note again that our choice of the DSF range for
$f_{esc}$ was motivated for the reasons outlined earlier, and
that other ranges for $f_{esc}$ are possible; the main point
that is demonstrated by Figure \ref{fig:fesca} is the power of
using any such band of independently known astrophysics to
constrain a cosmological parameter. Note also that $A$ would
have to be known to great accuracy to place any limits on
$f_{esc}$ that are stronger than the DSF band.

\begin{table}[t]
\caption{$V$ AND DIAGONAL ELEMENTS OF $W$ from $F = U W V^T$}
\begin{center}
\begin{tabular}{lcrrrrrl}
\tableline\tableline
$\sigma_8^2$ & & -0.3922 & 0.1470 & -0.5116 & 0.7498 & -0.0277 & 0.000 \\
$\Omega_b$ & & -0.5024 & -0.8592 & 0.0916 & -0.0320 & -0.0028 &
8.42E-16 \\ $h$ & & 0.5012 & -0.3673 & -0.7611 & -0.1856 & -0.0111
& 3.613E-15 \\ $n$ & & -0.5840 & 0.3242 & -0.3857 & -0.6342 &
-0.0545 & 1.599E-14 \\ $f_{esc}$ & & 0.0098 & -0.0039 & 0.0111 &
0.0041 & -0.2539 & 0.9671 \\ $f_\star$ & & 0.0373 & -0.0148 &
0.0420 & 0.0155 & -0.9652 & -0.2544 \\ & & & & & & \\ $w$ & &
18110.877 & 9251.715 & 330.074 & 12.455 & 0.0191 & 6.366E-16\\
\tableline
\end{tabular}

\tablecomments{Results of the Singular Value Decomposition of
$F_{6 \times 6}$ = $U W V^T$; $V$, weights $w$ for each column
in $V$, and parameters associated with each row in $V$ are
shown.}
\end{center}
\end{table}

So far, we have been using the Fisher matrix formalism for
specific pairs of parameters, while fixing the values of the
other parameters in the SM. The more general and proper way to
do this is to construct a 6 $\times$ 6 matrix for the parameter
set [$A$, $\Omega_b$, $h$, $n$, $f_{esc}$, $f_\star$], which
yields $\tau$. However, the analysis described in \S 3 implies
that only any five of these six parameters will be independent,
as the CMB data will determine the cosmological parameters and
$\tau$. Indeed, the 6 $\times$ 6 matrix, when constructed,
proves to be singular. We note here briefly some informative
aspects of performing singular value decomposition (SVD)
\citep{numrec} of $F_{6 \times 6}$, so that $F$ = $U W V^T$,
where $W$ is a diagonal $6 \times 6$ matrix containing the
singular values $w$. An element that has an anomalously low
value (close to zero) in $W$ implies that the corresponding
column in $V$ is a linear combination of parameters that will
not be well-constrained.

Table 1 shows the matrix $V$ with the corresponding column
weights $w$ resulting from this decomposition; also shown are
the parameters associated with each row in $V$, with $A$
expressed as $\sigma_8^2$. We see that the sixth weight is very
close to zero, so that the sixth column of $V$ contains
combinations of [$P_{\rm cosmo}, P_{\rm astro}$] that are poorly
constrained by this analysis. This column has terms
corresponding to essentially only $P_{\rm astro}$, the dominant
contribution coming from $f_{esc}$. This implies that $f_{esc}$
will not be well-determined from the CMB (via $\tau$), given the
reionization model considered here.  The first five columns of
$V$ also convey what combinations of these six parameters {\it
will} be constrained; we note that $f_{esc}$ has very small
contributions in these, i.e., to the information that can be
extracted from the CMB. In comparison, $f_\star$ can be better
determined from the CMB, as seen from columns 1--5, particularly
the fifth, where the dominant term is from $f_\star$. This
insensitivity of the CMB data to $f_{esc}$ can be traced back to
the stellar reionization model we adopted here; variations in
$f_\star$ affect $\tau$ more significantly than do those in
$f_{esc}$ (see Figures 12 and 15 in HL97).

The covariance matrix for $F$ can be found by, $\mathcal{C}
\equiv F^{-1} = V W^{-1} U^T$. As the ratio of $W$'s minimum to
maximum value, $\sim$ 3 $\times 10^{-20}$, is very small
compared with machine precision, we follow the usual technique
of adjusting the anomalously low singular value in $W$, here
$w_6$, to zero \citep{numrec}; despite this, the SVD inversion
of $F$ still produces an inaccurate covariance matrix, i.e.,
$\mathcal{C} \bullet F \neq I$.

\begin{figure}[ht]
\begin{center}
\includegraphics[height=3.5in]{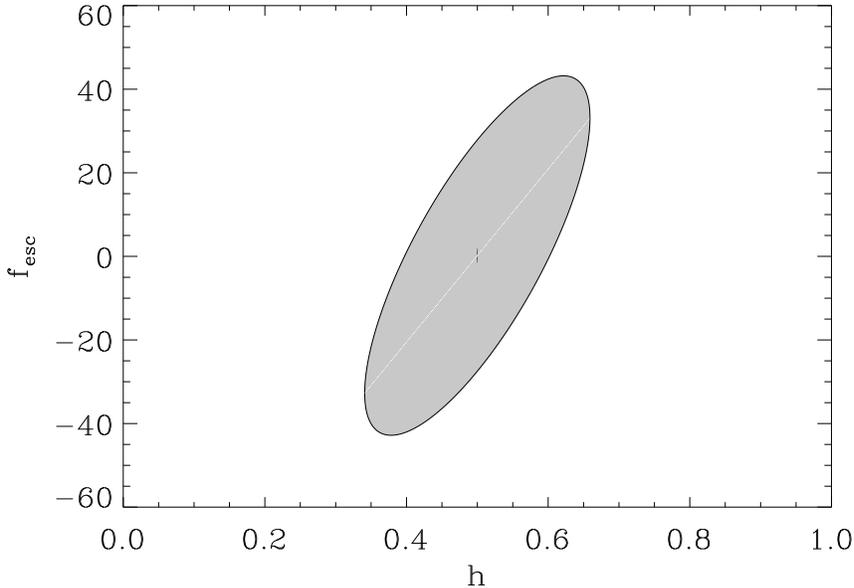}
\end{center}
\caption{Constraints from the CMB in the $f_{esc}$--$h$ plane;
larger ellipse represents current data, and smaller nested
ellipse (line) is from $PLANCK$.}
\label{fig:fesch}
\end{figure}
\begin{figure}[p]
\begin{center}
\includegraphics[height=3.5in]{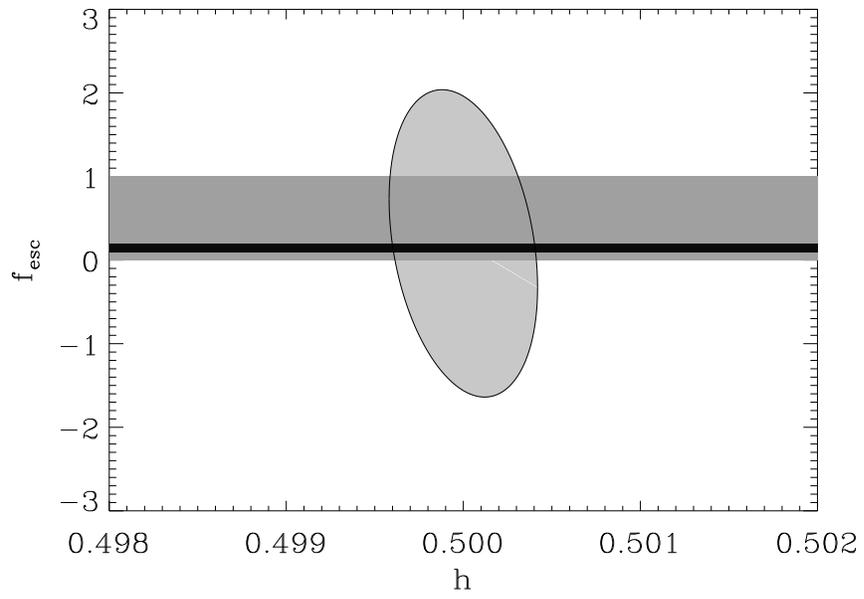}
\end{center}
\caption{Constraint from $PLANCK$ in the $f_{esc}$--$h$ plane;
light shaded band represents the entire allowed astrophysical
range of 0.1--1.0 for $f_{esc}$, and the dark shaded band
represents the permitted values of 0.1--0.2 for $f_{esc}$ from
DSF.}
\label{fig:feschpl}
\end{figure}

\begin{figure}[t]
\begin{center}
\includegraphics[height=3.5in]{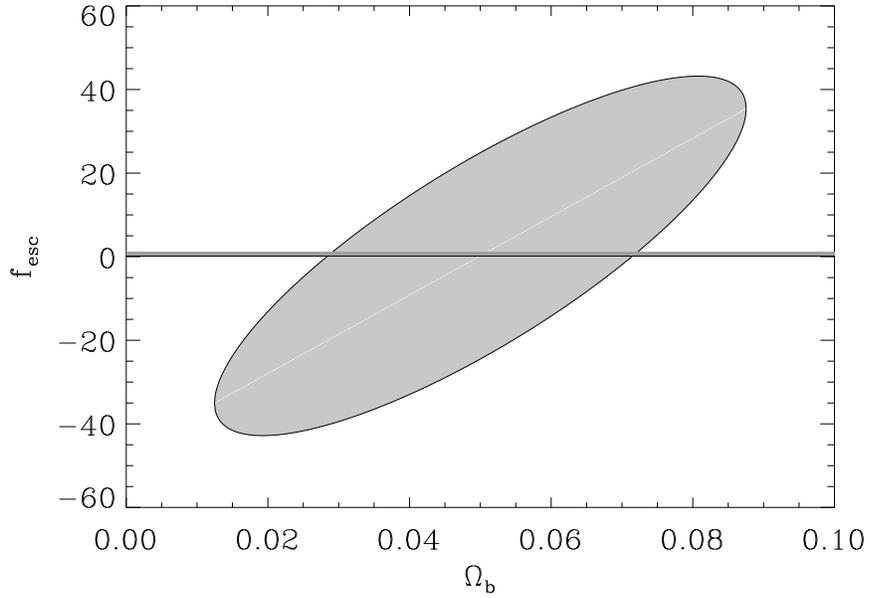}
\end{center}
\caption{Constraint from current CMB data in the
$f_{esc}$--$\Omega_b$ plane; shaded bands are the same as in
Figure \ref{fig:feschpl}.}
\label{fig:fescob}
\end{figure}

\begin{figure}[!hb]
\begin{center}
\includegraphics[height=3.5in]{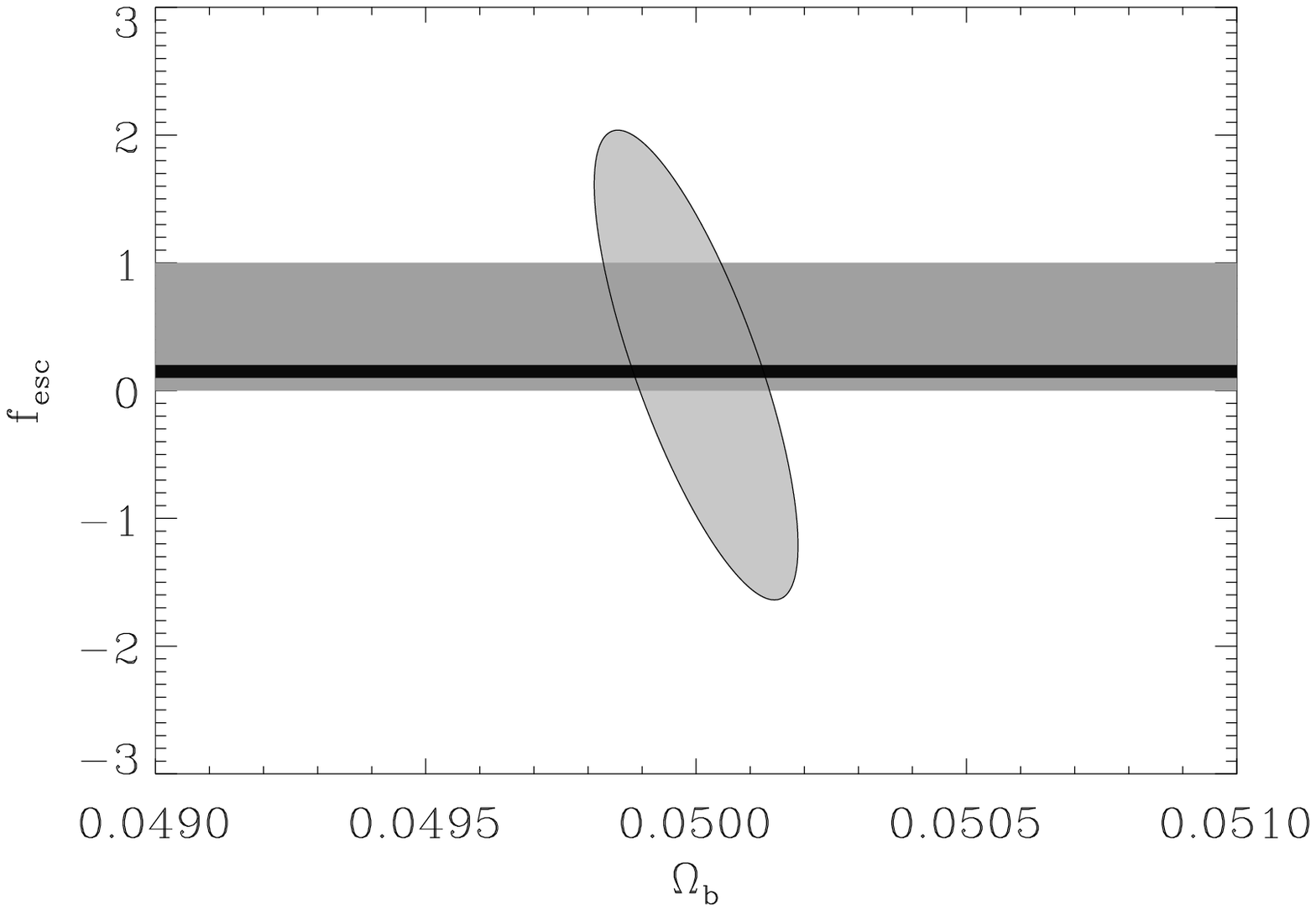}
\end{center}
\caption{Same as Figure \ref{fig:fescob}, but for data from
$PLANCK$.}
\label{fig:fescobpl}
\end{figure}

\begin{figure}[t]
\begin{center}
\includegraphics[height=3.5in]{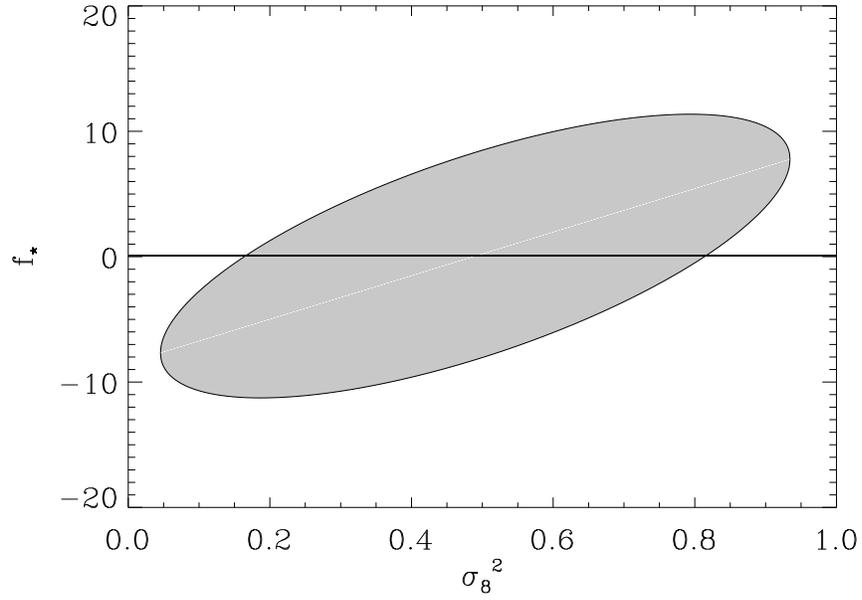}
\end{center}
\caption{Constraint from current CMB data in the
$f_\star$--$\sigma_8^2$ plane, where $\sigma_8^2 \propto A$;
shaded band represents the permitted astrophysical range of
0.01--0.15 for $f_\star$.}
\label{fig:fsa}
\end{figure}

\begin{figure}[hb]
\begin{center}
\includegraphics[height=3.5in]{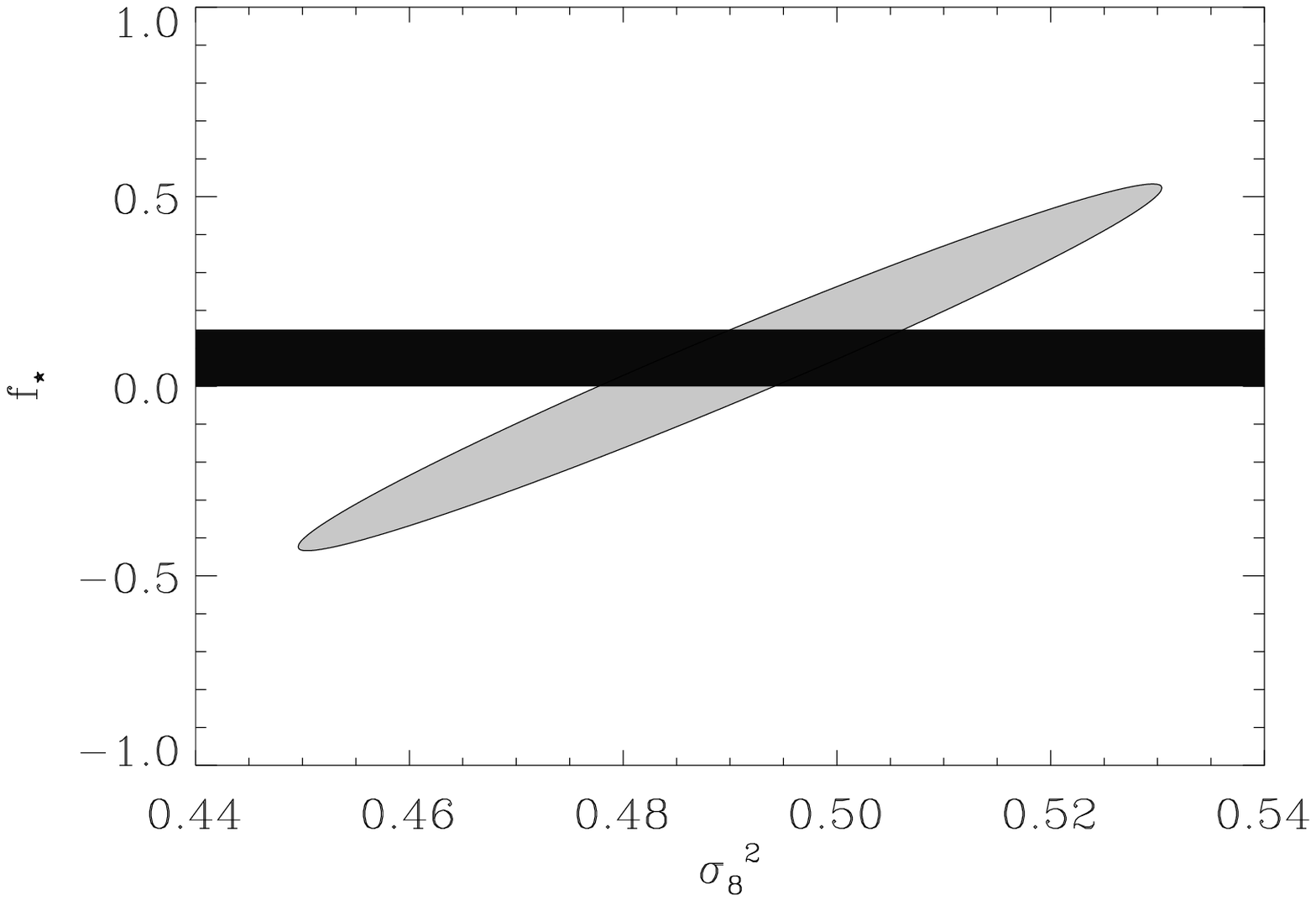}
\end{center}
\caption{Same as Figure \ref{fig:fsa}, but for data from
$PLANCK$.}
\label{fig:fsapl}
\end{figure}

\begin{figure}[htb]
\begin{center}
\includegraphics[height=3.5in]{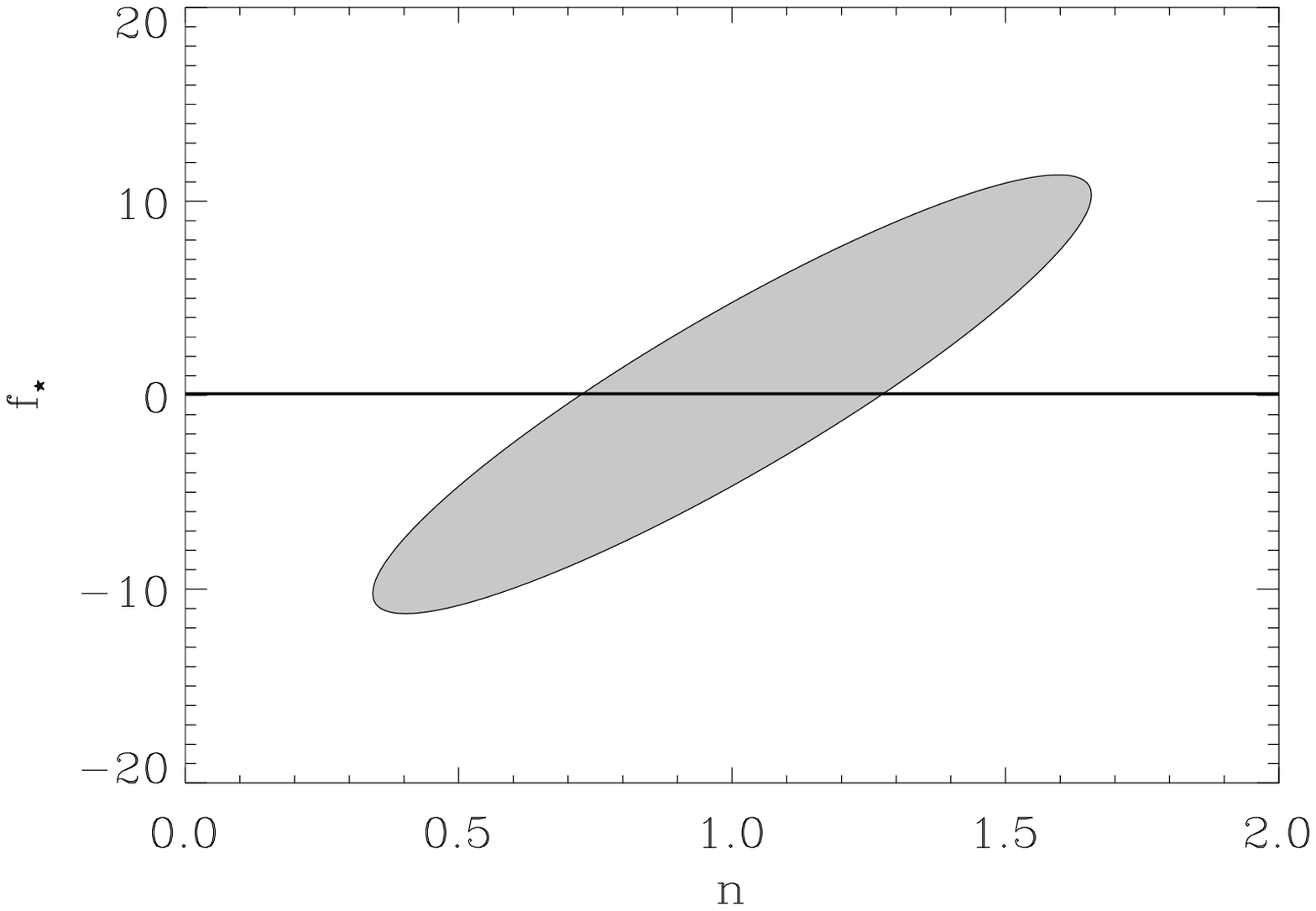}
\end{center}
\caption{Constraint from current CMB data in the $f_\star$--$n$
plane; shaded band is the same as in Figure \ref{fig:fsa}.}
\label{fig:fsn}
\end{figure}

\begin{figure}[htb]
\begin{center}
\includegraphics[height=3.5in]{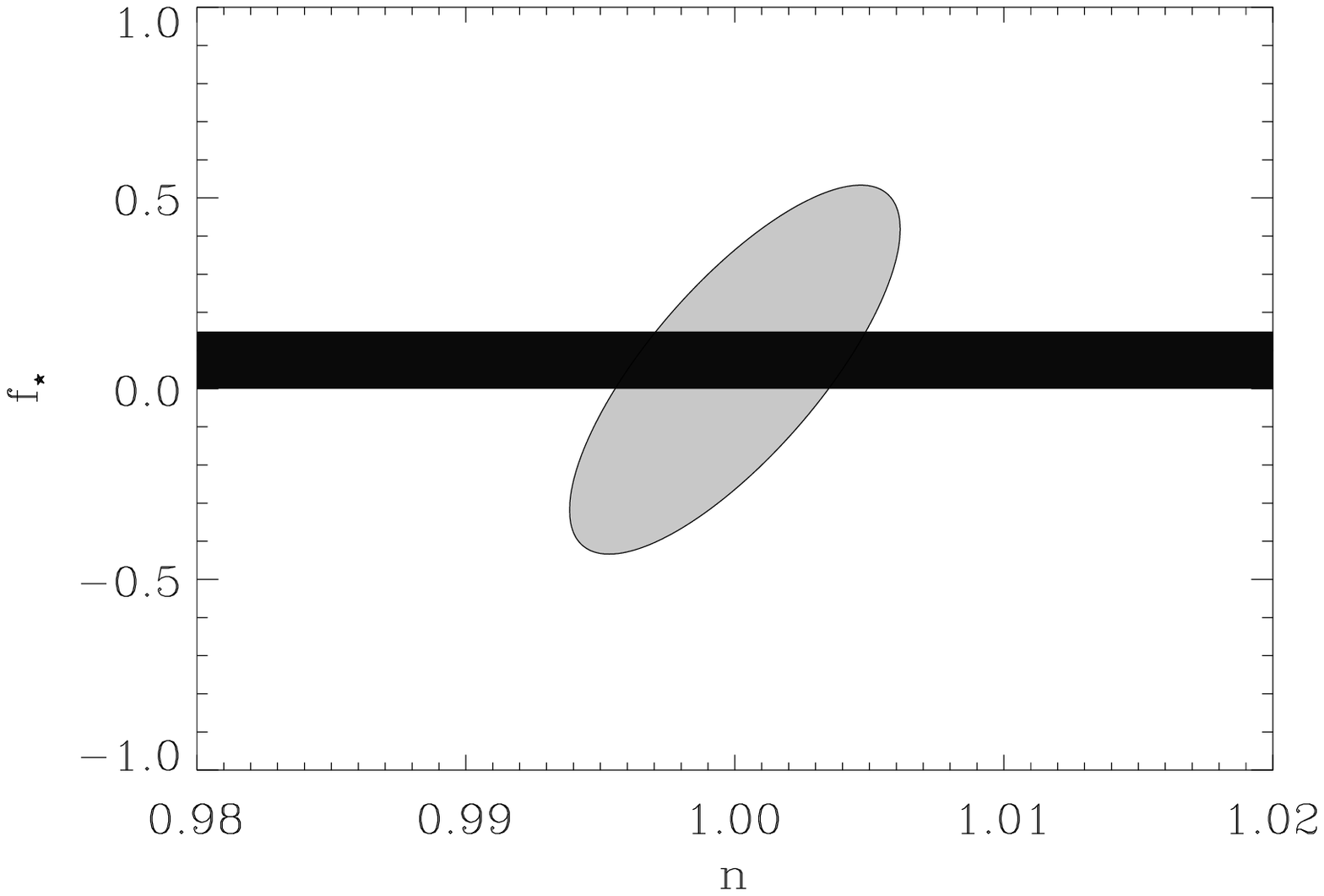}
\end{center}
\caption{Same as Figure \ref{fig:fsn}, but for data from
$PLANCK$.}
\label{fig:fsnpl}
\end{figure}

\begin{figure}[!t]
\begin{center}
\includegraphics[height=3.4in]{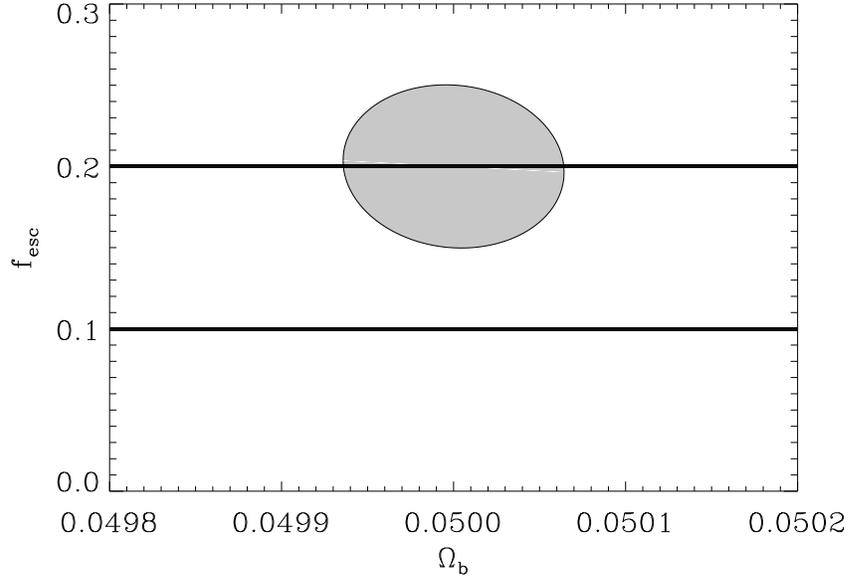}
\end{center}
\caption{Constraint from $PLANCK$ in the $f_{esc}$--$\Omega_b$
plane using temperature and polarization; lines represent the
allowed range for $f_{esc}$ from DSF.}
\label{fig:fescpol}
\end{figure}

\begin{figure}[!b]
\begin{center}
\includegraphics[height=3.4in]{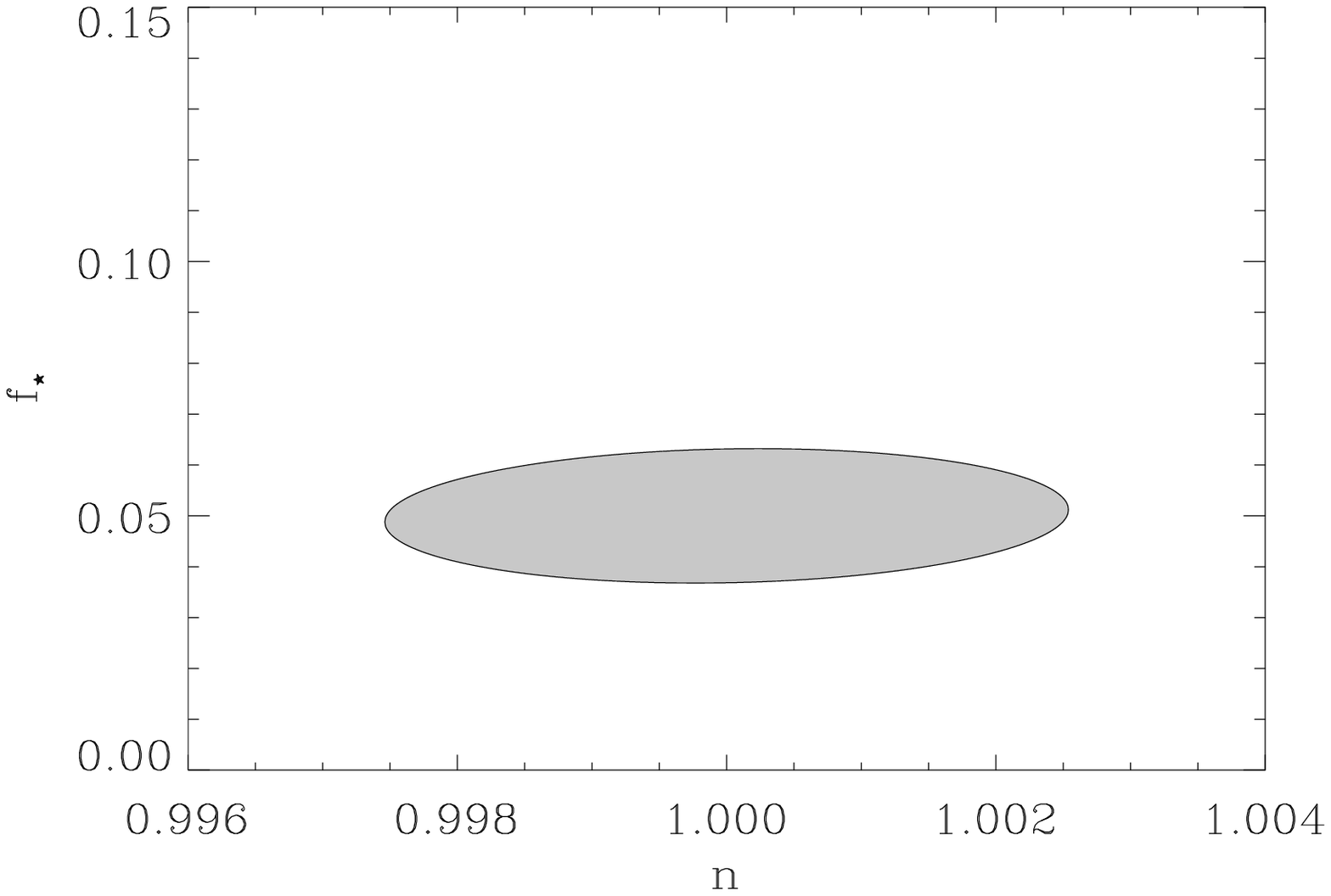}
\end{center}
\caption{Constraint from $PLANCK$ in the $f_\star$--$n$ plane
using temperature and polarization; the allowed astrophysical
range for $f_\star$ is the entire $y$-axis.}
\label{fig:fspol}
\end{figure}

We now proceed to work with the independent 5 $\times$ 5
subportions of the full 6 $\times$ 6 matrix, which translates to
$P_{\rm cosmo}$ and any one of $P_{\rm astro}$. These 5 $\times$
5 matrices are inverted, and the 2 $\times$ 2 submatrix of
interest is projected into the 2-parameter plane as the
appropriate error ellipse, which displays the confidence region
after marginalizing over the other parameters. The results of
this general Fisher matrix analysis are presented below for the
idealized specifications of current data and for those expected
from $PLANCK$ (\S 3). Only the temperature anisotropies from the
CMB are used for Figures \ref{fig:fesch}--\ref{fig:fsnpl}; the
polarization information expected from $PLANCK$ is included for
Figures \ref{fig:fescpol}--\ref{fig:fspol}. These plots are
intended to be examples of the constraints in various $P_{\rm
astro}$--$P_{\rm cosmo}$ subspaces.

Figure \ref{fig:fesch} shows the case of $f_{esc}$ vs. $h$; the
larger ellipse corresponds to current CMB data, and the nested
one (appearing as a tiny line) is from $PLANCK$. For all
subsequent cases, we show these ellipses separately; the
astrophysical range for $f_{esc}$ is omitted from this plot for
visual clarity. Figure \ref{fig:feschpl} shows the results
expected from $PLANCK$ alone for $f_{esc}$ vs. $h$, with the
light shaded horizontal band representing the full range of
$f_{esc}$ (0.1--1.0), and the dark band representing the DSF
values for $f_{esc}$ (0.1--0.2). The case of $f_{esc}$
vs. $\Omega_b$ is shown in Figures \ref{fig:fescob} and
\ref{fig:fescobpl}, for current CMB data and $PLANCK$
respectively, with the overplotted bands being the same as in
Figure \ref{fig:feschpl}.

Figures \ref{fig:fsa}--\ref{fig:fsapl} and
\ref{fig:fsn}--\ref{fig:fsnpl} display the respective cases of
$f_\star$ vs. $\sigma_8^2$ (where $\sigma_8^2$ $\propto A$), and
$f_\star$ vs. $n$. For these 4 plots, the horizontal dark band
represents the maximum astrophysical range of 0.01--0.15 for
$f_\star$; values below this range are unlikely to be sufficient
for reionization and values above this range must invoke IMFs
other than that of present-day galaxies in order to not violate
metal production or background light constraints.

We note here some generic features of Figures
\ref{fig:feschpl}--\ref{fig:fsnpl}. In all cases, the inclusion
of known constraints on the astrophysical parameters strengthens
the CMB's limits on cosmological parameters, even for the data
expected from $PLANCK$. This is particularly the case with
$f_\star$, due to $\tau$'s greater sensitivity to $f_\star$
relative to $f_{esc}$. Thus the 1-$\sigma$ error for $f_\star$
from the CMB is significantly smaller than that for $f_{esc}$
for all the cases shown here, making independent limits on the
former a more powerful complementary constraint for cosmological
parameters extracted from the CMB. As some illustrative examples
involving $PLANCK$ data, the {\it entire} astrophysical
permitted band for $f_\star$ reduces the 1-$\sigma$ error for
$\sigma_8$ from about 0.02 to less than 0.01 (Figure
\ref{fig:fsapl}), and for $n$, from 0.006 to $\sim$ 0.004. The
power to increase such constraints will only become better as
$f_\star$ (or $f_{esc}$) become better constrained themselves,
but it may not matter much for most cosmological parameters in
the post-$PLANCK$ era (as they will already be determined with
great precision), with the exception of $A$ or $\sigma_8$. For
this case, the method here may prove to be a valuable
cross-check.

Figures \ref{fig:feschpl}--\ref{fig:fsnpl} also reveal that even
the most promising cases of cosmological parameter determination
from the CMB's temperature information will not help to
constrain astrophysical parameters such as $f_\star$ or
$f_{esc}$, whose currently known ranges as shown through the
horizontal bands in each figure are typically much smaller than
what would be deduced from the joint confidence region. This is
partly due to the low value of $\tau$ itself of $\sim$ 0.06 in
our SM, which hinders its accurate determination from the CMB
data.


When polarization is included for the projected data from
$PLANCK$, we see, from the two examples shown in Figures
\ref{fig:fescpol}--\ref{fig:fspol}, that $f_{esc}$ can be
determined to about the same accuracy as the DSF allowed band,
but that the 1-$\sigma$ error for $f_\star$ is significantly
smaller than its astrophysical uncertainty. Thus, future CMB
data may be able to constrain the astrophysical aspects of
reionization. We recall, however, that we have neglected effects
from experimental noise or from foregrounds in our analysis,
which will enlarge the joint confidence regions in all the
figures. While this only strengthens the argument of the power
of astrophysical limits in constraining cosmology, the converse
situation, which appears hopeful from Figures
\ref{fig:fescpol}--\ref{fig:fspol}, is realistically tentative
at best. In short, it is possible in principle that the
astrophysics of a stellar reionization model can be constrained
by limits on cosmological parameters from $PLANCK$'s temperature
and polarization data, though this may prove difficult to
achieve.  We may simply have to await the view from $SIRTF$ and
the $NGST$ to determine the reionizing activity of the first
stars!

\clearpage
\section{Conclusions}

We have examined the power of a reionization model, given its
many cosmological and astrophysical parameters, to constrain
these input quantities when combined with parameter extraction
from the CMB. In the case of the well-known degeneracy between
$\tau$ and $A$ in their effects on the CMB, we have found that
this can be alleviated by the complementary information from a
reionization model, and that this remains a useful
cross-constraint even when allowing for the astrophysical
uncertainty in $\tau$.

When we eliminate $\tau$ and perform a more general Fisher
matrix analysis, we find that the astrophysical details of
reionization {\it can} be useful in further constraining the
CMB's limits on cosmological parameters, even in the case of the
expected temperature data from $PLANCK$. We have shown that
independent limits on the astrophysical inputs to reionization,
despite the current uncertainty in their values, reduce the
errors for cosmological parameters by a factor of at least
$\sim$ 2. Given that we have considered the most optimistic
parameter yield from CMB experiments (\S 3), the use of known
astrophysics can only become more valuable for realistic
experimental results. This is of particular value for $\sigma_8$
(or $A$) and $n$, given their implications for structure
formation and for theoretical models of the origin of the seeds
of structure in the early universe.

The converse situation-- using a projected exquisite
determination of a cosmological parameter to constrain
astrophysical reionization parameters-- does not yield quite as
interesting results with temperature data from current
experiments or from $PLANCK$, even though we made the most
optimistic assumptions; the 1-$\sigma$ errors for $f_{esc}$ or
$f_\star$ are larger than what are already known to be
reasonable. When the projected polarization data from $PLANCK$
is included, we found that $f_\star$ in particular may be
constrained to far greater accuracy than its current
astrophysical uncertainty; in practice, however, this may prove
difficult to achieve, given the effects of foregrounds and
instrumental noise which we have neglected here.

In summary, one may take away that the astrophysical details of
reionization can strengthen the limits on the cosmology of our
universe, beyond even the projected parameter yield from future
CMB data, and that there is more potential to a measurement of
$\tau$ than the determination of a single number out of a large
parameter space describing adiabatic CDM models.  These broad
conclusions are naturally subject to the assumptions made in
this analysis. The sizes of joint confidence regions derived
from the CMB data for any 2-parameter subspace is determined by
the full covariance matrix, whose elements' values are dependent
on the dimension of the chosen parameter space and the selected
parameters. The inclusion of more parameters has the generic
result of increasing the sizes of the error ellipses; therefore,
the primary results of this paper can only be strengthened when
parameter spaces larger than that analyzed here are considered.

In the sCDM cosmology assumed here, the values of $\tau$ in our
standard model were relatively low ($\sim 0.06$). In an open
universe, or one dominated by a cosmological constant
contribution, we expect larger average values of $\tau$ for a
fixed reionization model, as structures freeze out earlier,
resulting in a longer line-of-sight to the last scattering
surface at the reionization epoch. Increased $\tau$'s can also
result from higher values of $f_\star$ or $f_{esc}$, or from a
lower value of $M_C$ (\S 2), which would allow the first stars
to form earlier and more ubiquitously. As higher $\tau$'s will
have a better chance of being accurately determined from the
CMB, it would be interesting to analyze the constraints in this
paper, from both the reionization scenario and the CMB, for a
more general parameter space; we examine this in a forthcoming
work.

\acknowledgements

This work was presented as part of a dissertation to the
Department of Astronomy and Astrophysics at The University of
Chicago, in partial fulfillment of the requirements for the
Ph.D. degree.  I thank Scott Dodelson for his many valuable
suggestions towards this project, Angela Olinto for helpful
comments and Daniel Eisenstein for useful discussions. Parameter
estimation from the CMB was performed using CMBFAST by Uros
Seljak and Matias Zaldarriaga. This research was partly
supported by the NSF through the collaborative US-India project
No. INT-9605235, by NSF grant AST 94-20759, and by DOE grant
DE-FG0291 ER40606 at The University of Chicago.

\end{document}